\documentclass[11pt,a4paper]{article}

\usepackage{amsmath,amssymb}
\usepackage{graphicx}
\usepackage{booktabs}
\usepackage{geometry}
\usepackage{hyperref}
\usepackage{xcolor}
\title{Bayesian Optimization of Noisy Log-Likelihoods \\
Evaluated by Particle Filters 
-- One Parameter Case --}
\author{Genshiro Kitagawa\\[3mm]
Tokyo University of Marine Science and Technology\\
and\\
The Institute of Statistical Mathematics
}
\date{}

\begin{document}
\maketitle

\begin{abstract}
Likelihood functions evaluated using particle filters are typically noisy, computationally expensive, and non-differentiable due to Monte Carlo variability. These characteristics make conventional optimization methods difficult to apply directly or potentially unreliable. This paper investigates the use of Bayesian optimization for maximizing log-likelihood functions estimated by particle filters. By modeling the noisy log-likelihood surface with a Gaussian process surrogate and employing an acquisition function that balances exploration and exploitation, the proposed approach identifies the maximizer using a limited number of likelihood evaluations. Through numerical experiments, we demonstrate that Bayesian optimization provides robust and stable estimation in the presence of observation noise. The results suggest that Bayesian optimization is a promising alternative for likelihood maximization problems where exhaustive search or gradient-based methods are impractical. The estimation accuracy is quantitatively assessed using mean squared error metrics by comparison with the exact maximum likelihood solution obtained via the Kalman filter.
\end{abstract}

\section{Introduction}
Maximum likelihood estimation for state-space models is a fundamental problem
in time series analysis (\cite{AM 1979}), \cite{Kitagawa 2020}).
When the likelihood function is evaluated using particle filters,
Monte Carlo variability inevitably introduces noise into the likelihood surface,
making numerical optimization challenging (\cite{ADH 2010}).

Bayesian optimization provides a principled framework for optimizing
expensive and noisy objective functions (\cite{SSWAF 2016}).
However, its convergence behavior and estimation accuracy
for particle-filter-based likelihoods have not been sufficiently quantified,
mainly because the true maximum likelihood solution is usually unknown.

In this paper, we study Bayesian optimization of a noisy log-likelihood function
computed by a particle filter.
By considering a linear Gaussian state-space model with a single unknown parameter,
we are able to compute the exact maximum likelihood estimate using the Kalman filter.
This allows us to quantitatively evaluate the accuracy and uncertainty
of the estimated maximizer.

The main contributions of this paper are as follows:
(i) a systematic evaluation of Bayesian optimization with UCB
for noisy likelihood maximization,
(ii) quantitative error analysis based on repeated Monte Carlo experiments,
and (iii) illustrative examples of the posterior distribution
of the Gaussian process surrogate model.

\section{Problem Formulation}

\subsection{State-Space Model and State Estimation}
We consider a general state-space model for a time series $\{y_t\}$
consisting of a state equation and an observation equation,
\begin{eqnarray}
 x_t &=& f(x_{t-1}, v_{t-1}), \\
 y_t &=& h(x_t) + w_t,
\end{eqnarray}
where $x_t$ denotes the state vector and $f(x,v)$ and $h(x)$ are possibly 
nonlinear functions of $x$ and $v$ and $v$, respectively.
The process noise $v_t$ and the observation noise $w_t$
are mutually independent Gaussian random variables
with probability density functions $q(v)$ and $r(w)$, respectively.
The initial state $x_0$ is assumed to be distributed according to
the density function $p(x_0)$ (\cite{Kitagawa 1987}, \cite{Kitagawa 2020}).

Given a sequence of observations $y_{1:s} = \{y_1, \ldots, y_s\}$,
the posterior distribution of the state $x_t$ in a general state-space model,
$p(x_t \mid y_{1:s})$,
can be approximated using a sequential Monte Carlo method,
commonly referred to as a particle filter (\cite{GSS 1993}, \cite{Kitagawa 1996}, \cite{DFG 2001}).

The particle filter proceeds as follows:

\begin{itemize}
\item \textbf{Initialization:}
Draw particles $\{x_0^{(i)}\}_{i=1}^m$ independently from the prior distribution
$p(x_0)$ and assign equal weights $w_0^{(i)} = 1/m$.

\item \textbf{Prediction:}
For each particle $x_{t-1}^{(i)}$, generate a particle of the system noise
$v_t^{(i)} \sim r(v)$ and compute the predicted particle
\[
x_t^{(i)} \sim p(x_t \mid x_{t-1}^{(i)},v_t^{(i)}),
\]
according to the state transition model.

\item \textbf{Weight update:}
Update the particle weights using the observation likelihood,
\[
\tilde{w}_t^{(i)} = w_{t-1}^{(i)} \, p(y_t \mid x_t^{(i)}),
\]
and normalize them so that $\sum_{i=1}^m w_t^{(i)} = 1$.

\item \textbf{Resampling:}
If the effective sample size, $\mathrm{ESS}_t = \Bigl\{\sum_{i=1}^m \bigl(w_t^{(i)}\bigr)^2\Bigr\}^{-1}$ , falls below a predefined threshold,
resample the particles according to their weights
and reset the weights to $w_t^{(i)} = 1/m$.

\item \textbf{State estimation:}
Approximate the filtering distribution $p(x_t \mid y_{1:t})$
by the weighted particle set $\{x_t^{(i)}, w_t^{(i)}\}_{i=1}^m$.
\end{itemize}

\subsection{Log-likelihood Evaluation}

Using the particle filter, the filtering distribution
$p(x_t \mid y_{1:t})$ is approximated by a set of weighted particles
$\{x_t^{(i)}, w_t^{(i)}\}_{i=1}^m$.
Based on this approximation, the marginal likelihood of the observation
at time $t$ can be estimated as
\begin{equation}
\hat{p}(y_t \mid y_{1:t-1})
= \sum_{i=1}^m w_{t-1}^{(i)} \, p(y_t \mid x_t^{(i)}),
\end{equation}
where $m$ denotes the number of particles.

The likelihood of the entire observation sequence $y_{1:T}$
is given by the product of the predictive likelihoods,
\begin{equation}
p(y_{1:T}) = \prod_{t=1}^T p(y_t \mid y_{1:t-1}),
\end{equation}
which can be approximated using the particle filter as
\begin{equation}
\hat{p}(y_{1:T})
= \prod_{t=1}^T \hat{p}(y_t \mid y_{1:t-1}).
\end{equation}

Accordingly, the log-likelihood is estimated as
\begin{equation}
\hat{\ell}
= \sum_{t=1}^T \log \hat{p}(y_t \mid y_{1:t-1}).
\end{equation}
Due to the Monte Carlo nature of the particle filter,
this log-likelihood estimate is subject to random fluctuations,
and should therefore be regarded as a noisy function of the model parameters.
When emphasizing parameter dependence, the log-likelihood is denoted by
$\hat{\ell}(\theta)$.

The variance of this estimator depends on the number of particles
and increases as the particle count decreases (\cite{Kitagawa 2014}).

\subsection{Properties of Log-Likelihood Obtaine by Particle Filter}
An important property of the particle filter is that the estimator of the marginal likelihood is unbiased \cite{ADH 2010}.
Specifically, for a fixed parameter value, the particle filter provides an unbiased estimator of the likelihood $p(y_{1:T})$.
However, the corresponding log-likelihood estimator is generally biased due to Jensen's inequality and exhibits non-negligible Monte Carlo variability (\cite{Kitagawa 2014}).
This unbiasedness of the likelihood estimator holds regardless of the choice of proposal distribution and resampling scheme, provided that the particle filter is correctly implemented.

Although the likelihood estimator is unbiased,
its variance depends strongly on the number of particles.
Under mild regularity conditions, the variance of the log-likelihood estimator
decreases at an $O(1/N)$ rate as the number of particles increases (\cite{Kitagawa 2014}).
Conversely, when the number of particles is small,
the likelihood estimate exhibits substantial random fluctuations,
which manifest as noise in the objective function
for likelihood maximization.

In practice, this Monte Carlo variability makes direct optimization
of the log-likelihood difficult (\cite{ADH 2010}).
From the perspective of Bayesian optimization,
the estimated log-likelihood should therefore be regarded
as a noisy function whose noise variance decreases with increasing
particle count.
This observation motivates the use of Gaussian process surrogate models
with an explicit noise term to account for uncertainty
in likelihood evaluations.

\subsection{Normalization of Log-Likelihood}
Prior to Bayesian optimization, the estimated log-likelihood values
are normalized in order to improve numerical stability
and simplify the specification of Gaussian process hyperparameters.
Specifically, the log-likelihood is evaluated multiple times
at a small number of parameter values,
and the sample mean $\bar{\ell}$ and standard deviation $s_{\ell}$
are computed from these preliminary evaluations.

Each log-likelihood estimate $\hat{\ell}(\theta)$
is then standardized according to
\begin{equation}
\tilde{\ell}(\theta)
=
\frac{\hat{\ell}(\theta) - \bar{\ell}}{s_{\ell}},
\end{equation}
so that the transformed objective function
has approximately zero mean and unit variance.

This normalization removes the dependence of the optimization procedure
on the absolute scale of the log-likelihood
and facilitates the use of fixed Gaussian process hyperparameters,
such as the output scale and observation noise variance.
In particular, after standardization,
the observation noise variance can be set to unity,
as the effect of Monte Carlo variability
is implicitly absorbed into the normalized scale.

From the perspective of Bayesian optimization,
this procedure allows likelihood evaluations obtained
with different particle numbers 
to be treated in a consistent manner.
As a result, the optimization behavior becomes less sensitive
to the choice of tuning parameters
and more robust to stochastic fluctuations
in particle-filter-based likelihood estimates.

\subsection{An illustrative Example}
In the following numerical experiments, we consider an artificial time series
generated from the linear Gaussian state-space model
\begin{align}
  x_t &= x_{t-1} + v_t, \\
  y_t &= x_t + w_t ,
\end{align}
where $v_t \sim \mathcal{N}(0,\tau^2)$ and $w_t \sim \mathcal{N}(0,1.043)$
are mutually independent Gaussian noise processes,
and the initial state is given by $x_0 \sim \mathcal{N}(0,4)$.
We treat $\theta = \tau^2$ as an unknown parameter to be estimated (\cite{KG 1996},
\cite{Kitagawa 1996}, \cite{Kitagawa 2020}).

For this model, the predictive distribution
$p(y_t \mid y_{1:t-1})$ can be computed exactly using the Kalman filter,
which allows the log-likelihood $\ell(\theta)$
to be evaluated without numerical noise.
In contrast, when the log-likelihood is approximated using a particle filter,
Monte Carlo noise is inevitably introduced.

\begin{figure}[tbp]
\begin{center}
\includegraphics[width=160mm,angle=0,clip=]{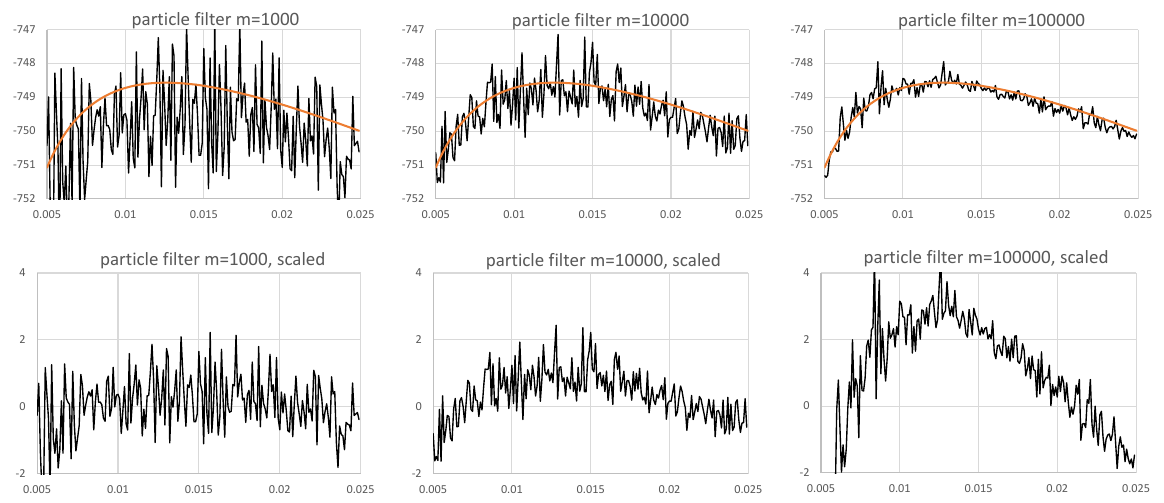}
\caption{Upper panels: Log-likelihood obtained by particle filters with the praticle number. $m$=1,000, 10,000 and 100,000, 0.005$\leq \tau^2 \leq$ 0.025. Red curve indicates the true log-likelihood function obtained by the Kalman filter. Lower panels: standardized log-likelihoods.}\label{Fig_Log-LK_particle-filter}
\end{center}
\end{figure}

The upper three panels of Figure \ref{Fig_Log-LK_particle-filter} show the log-likelihood
$\ell(\theta)$ evaluated by the particle filter
for $\theta \in [0.005,\,0.025]$
using $m = 10^4$, $10^5$, and $10^6$ particles.
The exact log-likelihood computed by the Kalman filter
is shown by the red curve.
When the number of particles is small (e.g., $m = 10^4$),
the particle-filter-based log-likelihood exhibits substantial fluctuations
around the true value.
As the number of particles increases,
these fluctuations become progressively smaller.

Table \ref{Tab_mean-variance-sd_of_log-likelihood} summarizes the results obtained by applying the particle filter
100 times with different random number seeds.
For each value of $\theta = 0.005 \times j$, $j = 1,\ldots,5$,
the mean, variance, and standard deviation of the estimated log-likelihood
are reported.
The exact log-likelihood values computed by the Kalman filter
are shown in the bottom row.
The results indicate that the log-likelihood estimates obtained
by the particle filter fluctuate around the true values,
and that their variance decreases as the number of particles increases.
Moreover, the variance strongly depends on the parameter value $\theta$;
in this example, larger values of $\theta$
lead to smaller variances in the log-likelihood estimates.

The lower panel of Figure \ref{Fig_Log-LK_particle-filter} shows the normalized log-likelihood,
obtained using the mean and standard deviation computed above
according to the normalization procedure described in the previous subsection.
After normalization, the log-likelihood fluctuates around zero,
and the magnitude of the fluctuations becomes relatively uniform
irrespective of the number of particles.

\begin{table}[tbp]
\caption{Mean, variance and standard deviation of the log-likelihood for various values of the parameter $\tau^2$ and the number of particles $m$.}
\label{Tab_mean-variance-sd_of_log-likelihood}
\begin{center}
\begin{tabular}{c|c|ccccc}
     &    &   \multicolumn{5}{c}{parameter value $\theta=\tau^2$} \\
\cline{3-7}
$m$ &   & 0.005 & 0.010 & 0.015 & 0.020 & 0.025  \\
\hline
       & mean & -752.68 & -749.63 & -749.26 & -749.99 & -750.76 \\
$10^4$ & var  & 1.7224 & 1.4055 & 1.0616 & 0.9096 & 0.7887 \\
       & sd   & 1.3124 & 1.1855 & 1.0303 & 0.9538 & 0.8881 \\
\hline
       & mean & -751.48 & -748.92 & -748.84 & -749.48 & -750.24 \\
$10^4$ & var  & 0.6565 & 0.3290 & 0.2013 & 0.1381 & 0.1379 \\
       & sd   & 0.8102 & 0.5736 & 0.4486 & 0.3716 & 0.3713 \\
\hline
       & mean & -750.94 & -748.68 & -748.73 & -749.36 & -750.19 \\
$10^5$ & var  & 0.1445 & 0.0477 & 0.0265 & 0.2821 & 0.0182 \\
       & sd   & 0.3802 & 0.2185 & 0.1629 & 0.1680 & 0.1348  \\
\hline
\multicolumn{2}{c|}{Kalman filter} & -751.08 & -748.72 & -748.66 & -749.22 & -750.01 \\
\hline
\end{tabular}
\end{center}
\end{table}
It should be noted, however, that as the number of particles $m$ increases,
the amplitude of variation of the underlying function becomes larger.
In the original (unnormalized) log-likelihood,
noise with a magnitude that strongly depends on the number of particles
is added to the same true log-likelihood function.
After normalization, this particle-number-dependent noise level
is largely equalized,
while the true underlying function to be estimated
effectively increases in scale as $m$ becomes larger.


\section{Bayesian Optimization with Noisy Likelihoods}

To maximize the log-likelihood estimated by the particle filter,
we employ Bayesian optimization, which is well suited
for objective functions that are expensive to evaluate and contaminated by noise (\cite{SSWAF 2016}, \cite{PWG 2013}).
In this framework, the estimated log-likelihood is treated
as a noisy function of the unknown parameter.

A Gaussian process (GP) is used as a surrogate model
for the log-likelihood function(\cite{RW 2006}).
The GP posterior provides both a mean estimate
and a measure of uncertainty at each parameter value,
allowing the optimization procedure to explicitly account
for Monte Carlo variability in likelihood evaluations.
An observation noise term is incorporated into the GP
to model the variance induced by the finite number of particles.

\subsection{Posterior of Gaussian Process}
In Bayesian optimization, the unknown objective function $f(x)$
is modeled as a Gaussian process (GP),
\[
f(x) \sim \mathcal{GP}\!\left(m(x), k(x,x')\right),
\]
where $m(x)$ and $k(x,x')$ denote the prior mean and covariance functions,
respectively.
Given a set of noisy observations
$\mathcal{D}_t = \{(x_i, y_i)\}_{i=1}^t$ with
\[
y_i = f(x_i) + \varepsilon_i, \qquad 
\varepsilon_i \sim \mathcal{N}(0,\sigma_n^2),
\]
the posterior distribution of $f(x)$ at an arbitrary input $x$
is again Gaussian.

Define the kernel matrix $\mathbf{K}_t \in \mathbb{R}^{t \times t}$ by
\[
(\mathbf{K}_t)_{ij} = k(x_i, x_j),
\]
and the kernel vector
\[
\mathbf{k}_t(x) = \bigl[k(x_1,x), \ldots, k(x_t,x)\bigr]^\top .
\]
Then the posterior predictive distribution of $f(x)$ is Gaussian,
\[
p\!\left(f(x)\mid \mathcal{D}_t\right)
=
\mathcal{N}\!\left(\mu_t(x),\, \sigma_t^2(x)\right),
\]
with posterior mean
\[
\mu_t(x)
=
\mathbf{k}_t(x)^\top
\left(\mathbf{K}_t + \sigma_n^2 \mathbf{I}\right)^{-1}
\mathbf{y}_t,
\]
and posterior variance
\[
\sigma_t^2(x)
=
k(x,x)
-
\mathbf{k}_t(x)^\top
\left(\mathbf{K}_t + \sigma_n^2 \mathbf{I}\right)^{-1}
\mathbf{k}_t(x),
\]
where $\mathbf{y}_t = [y_1,\ldots,y_t]^\top$ and $\mathbf{I}$ denotes
the identity matrix.


\subsection{Acqusition Function}
The posterior mean represents the current estimate of the objective function,
while the posterior variance quantifies the remaining uncertainty.
These two quantities are subsequently used to construct
the acquisition function that guides the selection of the next evaluation point.

At each iteration, the next parameter value to be evaluated
is selected by maximizing an acquisition function.
In this study, we adopt the upper confidence bound (UCB) defined as
\[
\mathrm{UCB}_t(x) = \mu_t(x) + \kappa_t s_t(x),
\]
where $\mu_t(x)$ and $s_t(x)$ denote the GP posterior mean
and standard deviation, $s_t(x) = \sqrt{\sigma_t^2(x)}$, respectively (\cite{SKKS 2010}).
UCB balances exploration and exploitation by combining
the posterior mean and standard deviation of the GP.
The exploration parameter $\kappa_t$ is allowed to vary with
the iteration index in order to balance exploration and exploitation.

Among various acquisition functions for Bayesian optimization,
the upper confidence bound (UCB) is particularly suitable
for maximizing log-likelihood functions estimated by particle filters.
The primary reason is that UCB explicitly accounts for both
the estimated function value and the associated uncertainty,
which is essential when likelihood evaluations are contaminated
by Monte Carlo noise.

The UCB acquisition function selects the next evaluation point
by maximizing a weighted sum of the posterior mean
and standard deviation of the Gaussian process surrogate model.
As a result, regions with high predictive uncertainty
are actively explored, preventing premature convergence
to spurious local optima caused by noisy likelihood estimates.
At the same time, regions with high posterior mean
are exploited as the optimization progresses.

Another advantage of UCB is its robustness to observation noise.
Unlike acquisition functions that rely on improvements
relative to the current best observed value,
UCB does not require repeated evaluations at the same parameter value
to reduce noise.
This property is particularly important in particle-filter-based
likelihood optimization, where repeated likelihood evaluations
are computationally expensive.

Furthermore, theoretical regret bounds are available for UCB
under mild assumptions, providing a principled justification
for its exploration--exploitation trade-off.
These properties make UCB a natural and effective choice
for Bayesian optimization of noisy log-likelihood functions.

\subsection{Iteration-Dependent Exploration Parameter $\kappa_t$}
In the UCB acquisition function, the exploration--exploitation balance
is controlled by the parameter $\kappa_t$.
In this study, $\kappa_t$ is allowed to depend on the iteration index $t$
in order to encourage exploration in the early stage
and gradually emphasize exploitation as uncertainty is reduced.

Specifically, we adopt the following iteration-dependent form:
\begin{equation}
\kappa_t = \sqrt{2 \log \left( \frac{t^2 \pi^2}{6 \delta} \right)},
\end{equation}
where $\delta \in (0,1)$ is a user-specified confidence parameter.
This choice is motivated by theoretical results on Gaussian process
upper confidence bounds, which guarantee sublinear cumulative regret
with high probability (\cite{SKKS 2010}).

The above formulation ensures that $\kappa_t$ increases slowly with $t$,
thereby maintaining sufficient exploration while avoiding excessive
sampling of highly uncertain but unpromising regions.
In the context of particle-filter-based likelihood optimization,
this behavior is particularly desirable,
as early exploration helps mitigate the effect of Monte Carlo noise,
while later exploitation enables precise localization
of the maximum likelihood estimate.

From a practical perspective, the parameter $\delta$
controls the overall exploration level.
Larger values of $\delta$ result in smaller $\kappa_t$,
leading to more aggressive exploitation,
whereas smaller values promote more conservative exploration.

\subsection{Convergence Criterion}
After evaluating the particle filter at the selected parameter,
the GP posterior is updated using the new noisy log-likelihood value.
Since the objective function evaluated by the particle filter
is contaminated by Monte Carlo noise,
convergence cannot be reliably assessed based solely
on improvements in observed log-likelihood values.
Instead, convergence is determined using uncertainty-based criteria
derived from the Gaussian process (GP) surrogate model.

At each iteration, the current estimate of the maximizer is defined as
\[
\hat{\theta}_t = \arg\max_{\theta} \mu_t(\theta),
\]
where $\mu_t(\theta)$ denotes the GP posterior mean.
Convergence of Bayesian optimization is declared
when both the change in the estimated maximizer
and the change in the normalized log-likelihood
fell below predefined thresholds.
Specifically, the stopping criteria were
$|x_{t+1}-x_t| < \epsilon_x$ and
$|\ell_{t+1}-\ell_t| < \epsilon_f$,
where $\epsilon_x$ was chosen relative to the kernel length scale
and $\epsilon_f$ was set according to the normalized noise level.
The thresholds $\varepsilon_x$ and $\varepsilon_f$ are chosen as
$\varepsilon_x = 0.01$ and $\varepsilon_f = 0.1$, which are commonly
used as a rule of thumb in practical Bayesian optimization settings.


Although convergence criteria were employed to terminate the optimization
in practical implementations,
all comparative evaluations in this study
were conducted using a fixed number of function evaluations.
This is because the convergence behavior depends on the choice of
kernel hyperparameters and stopping thresholds,
and therefore convergence time alone does not provide
a fair basis for comparison.

\subsection{Summary of Bayesian Optimization Procedure}
The Bayesian optimization procedure is summarized as follows:

\begin{itemize}
\item \textbf{Initialization:}
Select an initial set of parameter values
and evaluate the log-likelihood at these points
using the particle filter.

\item \textbf{Surrogate modeling:}
Construct a Gaussian process (GP) model
for the log-likelihood function,
treating the particle-filter estimates
as noisy observations.

\item \textbf{Acquisition function:}
Define an acquisition function based on the GP posterior.
In this study, the upper confidence bound (UCB)
is employed to balance exploration and exploitation.

\item \textbf{Parameter selection:}
Determine the next parameter value
by maximizing the acquisition function
using a numerical optimization method.

\item \textbf{Likelihood evaluation:}
Evaluate the particle filter at the selected parameter value
to obtain a new noisy estimate of the log-likelihood.

\item \textbf{Model update:}
Update the GP posterior with the newly obtained observation.

\item \textbf{Iteration and stopping:}
Repeat the above steps until a convergence criterion
based on uncertainty measures of the estimated maximizer is satisfied.
\end{itemize}

Notes: 
Prior to GP modeling, the log-likelihood values
are standardized to remove scale dependence.
The observation noise variance in the GP decreases
as the number of particles increases.
When the likelihood is available in closed form,
such as in linear Gaussian models,
Bayesian optimization reduces to deterministic function maximization.
At each iteration, the maximizer of the UCB is searched for
using a hybrid strategy combining grid search and Brent's method.
The particle filter is then executed at the selected parameter value,
and the GP posterior is updated accordingly.

\section{Numerical Experiments}

\subsection{Experimental Setup}
In this section, we consider the problem of estimating the variance of
the system noise in the one-dimensional linear Gaussian state-space
model introduced in Section~2.5.
All experiments are conducted over a fixed parameter range
$0.005 \leq \tau^2 \leq 0.025$.

The number of particles used in the particle filter is set to
$m = 1{,}000$, $10{,}000$, and $100{,}000$.
Since the log-likelihood is normalized in advance, the signal variance
of the Gaussian process kernel is fixed at $\sigma_f = 1$.
We examine four values of the noise variance parameter,
$\sigma_n = 0.2, 0.3, 0.5,$ and $1.0$, and five values of the length-scale
parameter, $\ell = 0.1, 0.2, 0.3, 0.5,$ and $1.0$.

Although the normalization of the log-likelihood suggests that setting
$\sigma_n = 1$ alone might be sufficient, Table~1 shows that the variance
of the log-likelihood estimates varies substantially across the
parameter range.
For this reason, the normalization is performed using the maximum
variance over the parameter domain, and several values of $\sigma_n$
not exceeding unity are considered.

All computations are repeated 100 times using independent random seeds,
and the average behavior over these runs is analyzed.

\subsection{Evaluation Criteria}
To evaluate the estimation accuracy of Bayesian optimization,
the particle filter is executed 100 times using independent random seeds.
The optimization is run for 100 iterations, and for each iteration
$i = 0, \ldots, 100$, the mean squared errors (MSEs) of the estimated
maximizer and the estimated maximum log-likelihood are computed as
\begin{align}
  \mathrm{MSE}(x_i)
  &= \mathbb{E}\!\left[ (\hat{\theta}_i - \theta^*)^2 \right], \\
  \mathrm{MSE}(\ell_i)
  &= \mathbb{E}\!\left[ \bigl( \ell(\hat{\theta}_i) - \ell(\theta^*) \bigr)^2 \right].
\end{align}
Here, $\theta^*$ denotes the maximum likelihood estimate of $\theta$
obtained using the Kalman filter and $\hat{\theta}_i$ denotes the estimated
maximize at the $i$-th iteration.


\begin{figure}[tbp]
\begin{center}
\includegraphics[width=160mm,angle=0,clip=]{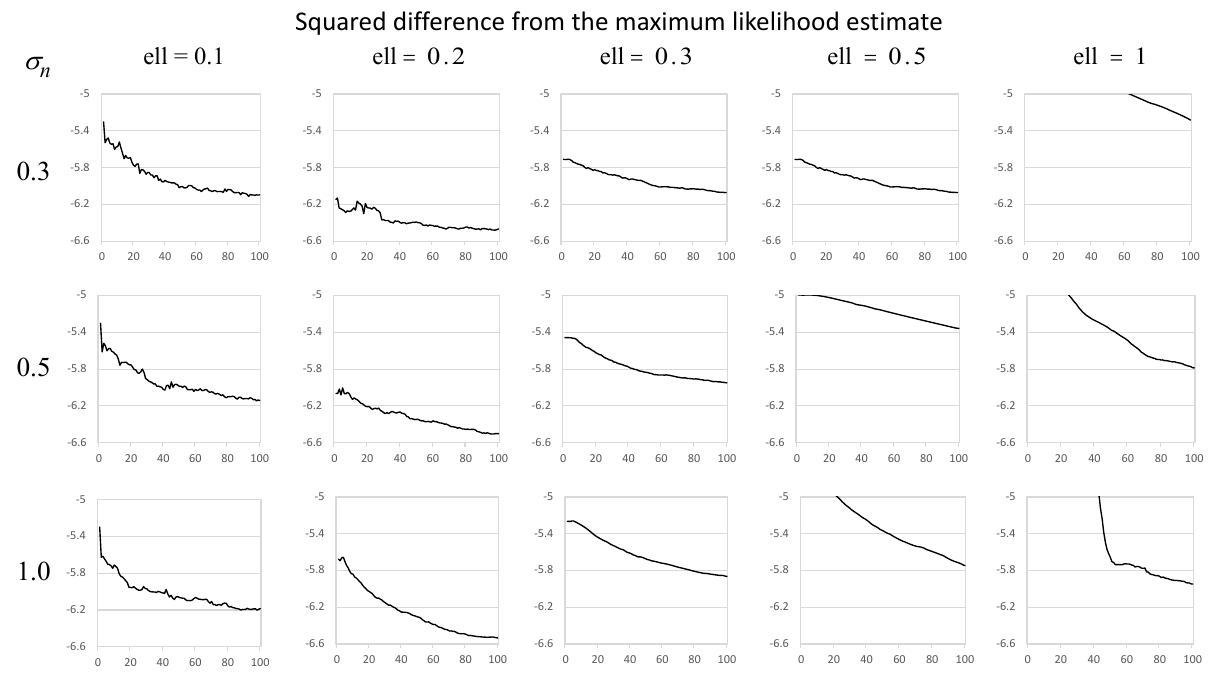}
\includegraphics[width=160mm,angle=0,clip=]{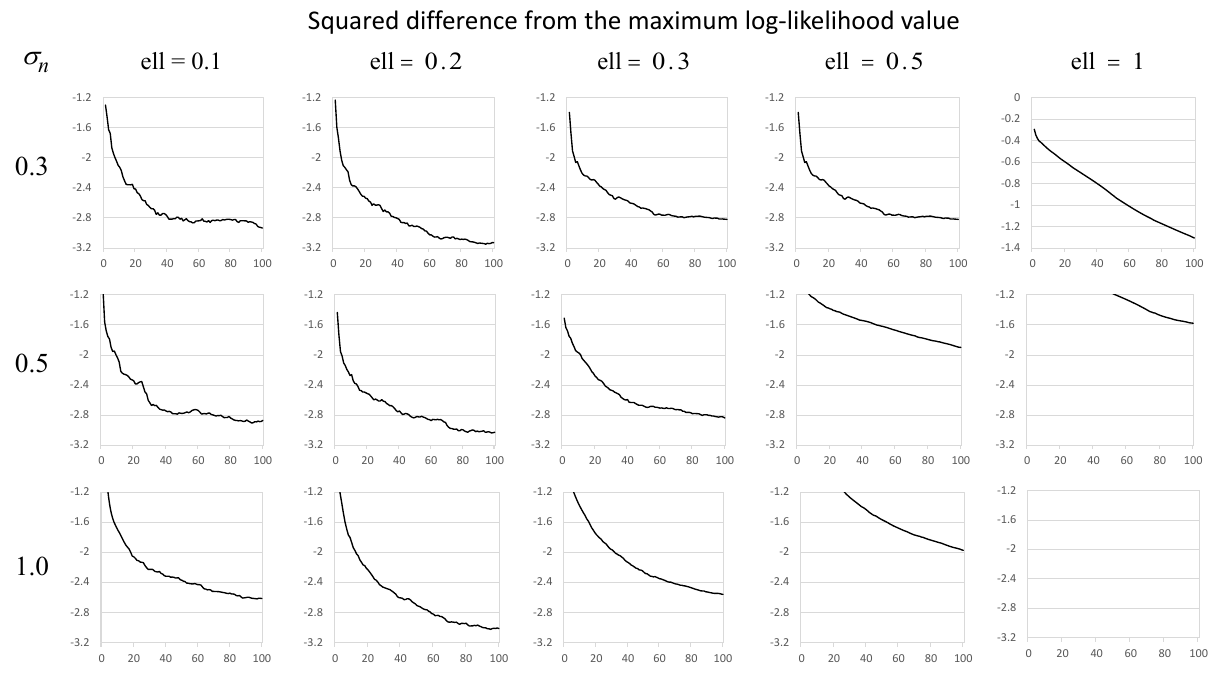}
\caption{Squared error of Bayesian optimization of log-likelihood computed by particle filter obtained by 100 repetition of optimization. Upper panels: $\log_{10}$MSE($x_i$) versus iter, lower panels: $\log_{10}$MSE($\ell(x_i)$) versus iter.}\label{Fig_squared error of BO-UCB}
\end{center}
\end{figure}

The mean squared error (MSE) is adopted as the evaluation metric for the
estimation accuracy for the following reasons.

First, MSE simultaneously accounts for both bias and variance of the estimator.
In the present problem, the log-likelihood evaluated by the particle filter
contains Monte Carlo noise, and Bayesian optimization may converge to a biased
estimate when the noise level is large or the number of iterations is limited.
MSE provides a unified measure that penalizes both systematic deviation from
the true maximizer and random fluctuations caused by stochastic likelihood
evaluation.

Second, since the true maximizer $\theta^*$ and the corresponding maximum
log-likelihood $\ell(\theta^*)$ can be computed exactly using the Kalman filter
for the considered linear Gaussian state-space model, MSE allows for a direct
and quantitative comparison between the estimated and true values.
This makes MSE a natural and interpretable criterion for performance evaluation.

Third, MSE is well suited for averaging over repeated experiments with different
random seeds.
By computing the expectation over multiple independent runs, MSE captures the
average convergence behavior of Bayesian optimization under stochastic
likelihood evaluations, rather than the outcome of a single realization.

Finally, evaluating MSE as a function of the iteration index enables a clear
assessment of convergence speed and stability.
A rapid decrease in MSE indicates fast convergence toward the true optimum,
while a plateau or slow decay reflects limitations imposed by noise or
hyperparameter choices.

The upper 15 panels of Figure \ref{Fig_squared error of BO-UCB} show how the logarithm of the mean squared error of
the estimated maximizer, $\log_{10}\mathrm{MSE}(x_i)$, evolves as a function of
the iteration index for five values of $\ell$ and three values of $\sigma_n$.
When $\ell = 0.2$, the estimation error is small from the early stages of the
optimization and remains smaller than those obtained with other values of
$\ell$ even at $\text{iter}=100$.
The effect of $\sigma_n$ is relatively minor; however, a closer inspection
reveals that $\sigma_n = 0.3$ yields a better approximation of the maximizer in
the early iterations, whereas $\sigma_n = 1$ provides slightly better accuracy
in the final stage of the optimization.

The lower panels of Figure \ref{Fig_squared error of BO-UCB} display the logarithm of the mean squared error of the
estimated log-likelihood values.
Again, $\ell = 0.2$ achieves the highest overall accuracy.
Moreover, relatively good performance is observed for the parameter
combinations $\sigma_n = 0.3$ or $0.5$ with $\ell = 0.1$, $0.2$, or $0.3$.
This behavior suggests that the log-likelihood function is relatively flat near
its maximum, so that moderate deviations in the estimated maximizer result in
only small changes in the log-likelihood value.

Table \ref{Tab_log-squared_errors_of_x_and_f(x)} reports the numerical values of $\log_{10}\mathrm{MSE}(x_i)$ and
$\log_{10}\mathrm{MSE}(\ell(x_i))$ for $\text{iter}=10$ and $100$.
For reference, additional results for $\sigma_n = 0.2$ and $2.0$ are also
included.
Bold red numbers indicate the minimum values for each case, while red numbers
in normal weight denote results whose difference from the minimum is within
$0.30$, corresponding to an error variance no more than twice the minimum.
Blue numbers indicate values whose difference from the minimum exceeds $1.0$,
i.e., cases where the error variance is at least ten times larger than the
minimum.

From this table, it is again evident that $\ell = 0.2$ provides the most accurate
estimation of the maximizer.
At $\text{iter}=10$, the minimum error is achieved with $\sigma_n = 0.3$, whereas
at $\text{iter}=100$, $\sigma_n = 1.0$ yields the smallest error.
However, for $\text{iter}=100$, the differences among various $\sigma_n$ values
are not pronounced, indicating that the final accuracy is relatively insensitive
to $\sigma_n$.

Regarding the estimation error of the maximum log-likelihood value, good accuracy
is obtained for both $\text{iter}=10$ and $100$ when
$\ell \in \{0.1, 0.2, 0.3\}$ and $\sigma_n \in \{0.2, 0.3, 0.5\}$.
This observation further supports the notion that the log-likelihood surface is
relatively flat in the vicinity of its maximum.

Overall, these results indicate that, at least for the present numerical example,
the parameter choices $\ell = 0.2$ and $\sigma_n \in \{0.3, 1.0\}$ yield favorable
estimation accuracy.

\begin{table}[tbp]
\caption{Logarithm of squared errors of $x$ and $f(x)$, $\log_{10}$MSE($x_i$), $\log_{10}$MSE($\ell(x_i)$).}
\label{Tab_log-squared_errors_of_x_and_f(x)}
\begin{center}
\begin{tabular}{c|c|ccccc||ccccc}
     &    &   \multicolumn{5}{c||}{$\log_{10}$MSE($x_i$)} & \multicolumn{5}{|c}{$\log_{10}$MSE($\ell(x_i)$)} \\
\cline{3-12}
     &    &   \multicolumn{5}{c||}{$\ell$} & \multicolumn{5}{|c}{$\ell$} \\
iter &$\sigma_n$ & 0.1 & 0.2 & 0.3 & 0.5 & 1 & 
                   0.1 & 0.2 & 0.3 & 0.5 & 1 \\
\hline
   &0.2& -5.54 & \color{red}-6.10 & \color{red}-5.98 & -5.33 & \color{cyan}-4.70 & \color{red}-2.01 & \color{red}-2.25 & \color{red}\textbf{-2.36} & -2.03 & \color{cyan}-0.86 \\
   &0.3& -5.57 & \color{red}\textbf{-6.27} & -5.76 & \color{cyan}-5.16 & \color{cyan}-4.48 & \color{red}-2.13 & \color{red}-2.30 & \color{red}-2.23 & -1.64 & \color{cyan}-0.49 \\
10 &0.5& -5.64 & \color{red}-6.10 & -5.51 & \color{cyan}-5.00 & \color{cyan}-4.57 & \color{red}-2.04 & \color{red}-2.26 & -1.97 & \color{cyan}-1.24 & \color{cyan}-0.32 \\
   &1.0& -5.71 & -5.84 & -5.31 & \color{cyan}-4.85 & \color{cyan}-3.83 & -1.70 & -1.87 & -1.41 & \color{cyan}-0.78 &  \color{cyan}0.21 \\
   &2.0& -5.95 & -5.65 & \color{cyan}-5.23 & \color{cyan}-4.92 & \color{cyan}-3.81 & \color{cyan}-0.93 & \color{cyan}-0.95 & \color{cyan}-0.78 & \color{cyan}-0.39 &  \color{cyan}-0.20 \\
\hline
   &0.2& -6.05 & \color{red}-6.38 & -6.20 & \color{cyan}-5.44 & \color{cyan}-4.78 & \color{red}-2.93 & \color{red}-2.91 & \color{red}-2.86 & -2.15 & \color{cyan}-1.09 \\
   &0.3& -6.10 & \color{red}-6.46 & -6.07 & \color{cyan}-5.31 & \color{cyan}-5.28 & \color{red}-2.93 & \color{red}\textbf{-3.13} & -2.82 & \color{cyan}-1.93 & \color{cyan}-1.31 \\
100&0.5& -6.14 & \color{red}-6.50 & -5.95 & \color{cyan}-5.36 & -5.79 & \color{red}-2.87 & \color{red}-3.03 & \color{red}-2.84 & \color{cyan}-1.90 & \color{cyan}-1.58 \\
   &1.0& -6.19 & \color{red}\textbf{-6.53} & -5.86 & -5.75 & -5.95 & -2.61 & \color{red}-3.01 & -2.56 & \color{cyan}-1.97 & \color{cyan}-0.91 \\
   &2.0& \color{red}-6.52 & \color{red}-6.44 & -5.97 & \color{cyan}-5.51 & \color{cyan}-3.81 & -2.20 & -2.50 & -2.20 & \color{cyan}-1.40 &  \color{cyan}0.32 \\
\hline
\end{tabular}
\end{center}
\end{table}


\section{Posterior Distribution Example and Convergence Process}
In the following, we provide additional insight into the Bayesian optimization
of the log-likelihood evaluated using a particle filter.
Specifically, we present representative examples of the Gaussian process
posterior distribution, together with illustrative examples of the convergence
behavior of the incremental changes in the estimated maximizer and the
estimated maximum log-likelihood value.

\subsection{Posterior Distribution Example}

\begin{figure}[tbp]
\begin{center}
\includegraphics[width=160mm,angle=0,clip=]{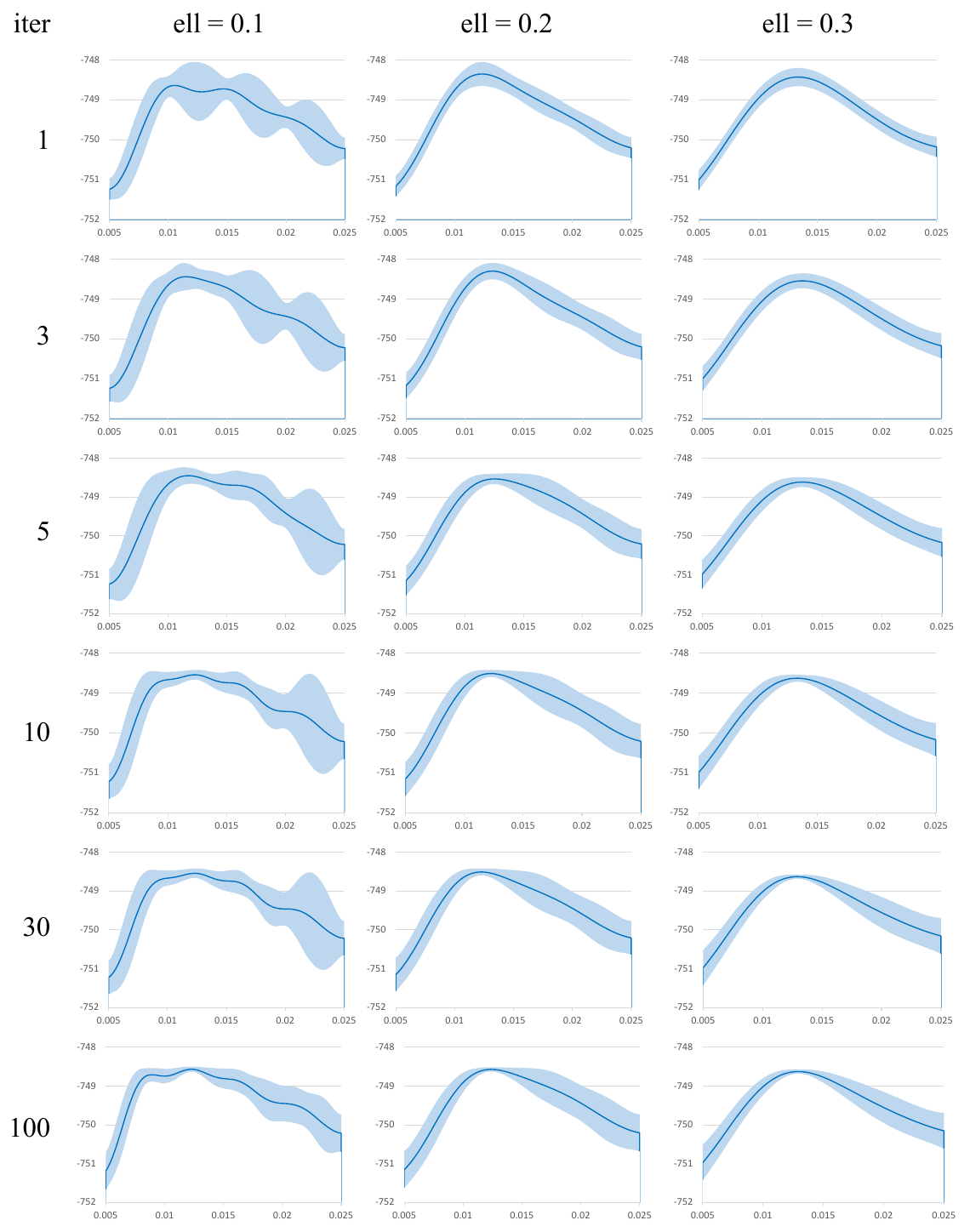}
\caption{Posterior distribution of Bayesian process. iter=1, 3, 5, 10, 30 and 100. $\sigma_f=1$, $\sigma_n$=0.3, $\ell$=0.1, 0.2 and 0.3.}\label{Fig_Bayesian process posterior distribution}
\end{center}
\end{figure}

Figure \ref{Fig_Bayesian process posterior distribution} illustrates the evolution of the posterior distribution of the
Gaussian-process surrogate model, obtained by fixing $\sigma_n = 0.3$ and varying
$\ell$ among $0.1$, $0.2$, and $0.3$.
As an initial training phase, the log-likelihood is evaluated using the particle
filter at five locations,
$x = 0.005, 0.010, 0.015, 0.020$, and $0.025$, after which Bayesian optimization is
performed.
The figure shows the posterior distributions of the surrogate model at six
iterations, $\text{iter} = 1, 3, 5, 10, 30$, and $100$.

In each panel, the solid blue curve represents the posterior mean $\mu(x)$,
while the light blue shaded region indicates the interval
$\mu(x) \pm \kappa_{\text{iter}} \, s(x)$.
The maximizer of $\mu(x)$ corresponds to the current estimate of the maximum
log-likelihood, and the associated $x$ value is taken as the estimated maximizer.
In contrast, the maximizer of $\mu(x) + \kappa_{\text{iter}} \, s(x)$ determines the next
sampling location, at which the log-likelihood is evaluated and the Gaussian
process posterior is subsequently updated.

For $\ell = 0.1$, the uncertainty bands are extremely wide in regions without
observations, leading to exploration over a broad range of the parameter space.
In contrast, when $\ell = 0.3$, the uncertainty band is nearly uniform during
the early iterations, but as the number of iterations increases, the uncertainty
shrinks only in the central region.
This behavior indicates that the exploration gradually concentrates around the
central area of the parameter space.

In all three cases, accurate estimates of both the maximizer and the maximum
log-likelihood are obtained.
However, it should be noted that the surrogate model represented by the Gaussian
process posterior does not necessarily provide a faithful reconstruction of the
true log-likelihood function, shown as the red curve in Figure \ref{Fig_Log-LK_particle-filter}.

\subsection{Convergence of Bayesian Optimization Process}
\begin{figure}[tbp]
\begin{center}
\includegraphics[width=160mm,angle=0,clip=]{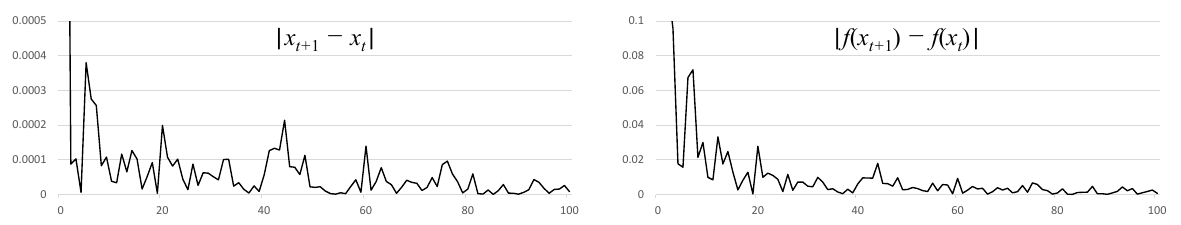}
\caption{Convergence of $|x_{t+1}-x_t|$ and $|\ell(x_{t+1})-\ell(x_t)|$, Bayesian process parameters are $\sigma_f$=1, $\sigma_n$=0.3, $\ell$=0.2 and the number of particle $m$ is 100,000.}\label{Fig_convergence of BO}
\end{center}
\end{figure}

Figure \ref{Fig_convergence of BO} illustrates the evolution of the absolute changes in the estimated
maximizer, $|x_{t+1} - x_t|$, and in the estimated log-likelihood value,
$|\ell(x_{t+1}) - \ell(x_t)|$, as the iteration progresses.
The Bayesian optimization hyperparameters are set to $\sigma_f = 1$,
$\sigma_n = 0.3$, and $\ell = 0.2$, and the number of particles in the particle
filter is fixed at $100{,}000$.
Both quantities exhibit fluctuations in the early stages of the optimization
but gradually approach zero as the number of iterations increases.

As a convergence criterion, we require that 
$|x_{t+1} - x_t| < 0.01$ and $|\ell(x_{t+1}) - \ell(x_t)| < 0.1$ hold for several
consecutive iterations.
It is evident from the figure that both conditions are satisfied relatively
early in the optimization process, suggesting stabilization of the estimated
maximizer and objective value.
However, these criteria are heuristic and are intended to indicate practical
termination rather than optimal statistical accuracy.
In particular, as Table \ref{Tab_log-squared_errors_of_x_and_f(x)} indicates, the mean squared errors of both the maximizer and the
log-likelihood estimates may continue to decrease with additional iterations,
as demonstrated by the results at later iterations.

\section{Discussion}

The numerical results demonstrate that Bayesian optimization with a
Gaussian-process surrogate model and the UCB acquisition function can
successfully maximize noisy log-likelihood functions evaluated by a particle
filter.
Even though the surrogate model does not necessarily reconstruct the true
log-likelihood function over the entire parameter space, accurate estimates of
both the maximizer and the maximum log-likelihood value are obtained.

The role of the Gaussian-process hyperparameters is also clarified.
The length-scale parameter $\ell$ primarily controls the spatial extent of
information propagation and strongly influences the convergence speed in the
early iterations.
Smaller values of $\ell$ lead to faster localization near the optimum, whereas
moderate values yield better stability in later stages.
The noise parameter $\sigma_n$ affects the balance between exploration and
exploitation: smaller values favor early convergence, while larger values
provide slightly better final accuracy.
After normalization of the log-likelihood, the influence of $\sigma_f$ is found
to be limited.

An important observation is that the estimation accuracy of the maximizer and
that of the maximum log-likelihood value are not necessarily equivalent.
When the log-likelihood surface is relatively flat around its maximum, moderate
errors in the estimated maximizer may result in only small errors in the
estimated maximum value.

The convergence criterion based on the incremental changes
$|x_{t+1}-x_t|$ and $|f(x_{t+1})-f(x_t)|$ is shown to be effective and robust to
stochastic fluctuations.
However, it should be noted that convergence in terms of these criteria does not
guarantee convergence to the true optimum, particularly when the likelihood
surface is multimodal.

Finally, the present study is limited to a one-dimensional linear Gaussian
state-space model.
Extensions to higher-dimensional and nonlinear models remain an important topic
for future research.

\section{Conclusion}

This paper investigated Bayesian optimization for maximizing noisy
log-likelihood functions evaluated by a particle filter.
Using a linear Gaussian state-space model for which the true maximum likelihood
estimate is available, we systematically evaluated the estimation accuracy and
convergence behavior of the optimization process.

The results indicate that Bayesian optimization with appropriate normalization
and hyperparameter settings can reliably estimate both the maximizer and the
maximum log-likelihood value, even in the presence of substantial Monte Carlo
noise.
The proposed evaluation framework, based on mean squared error and
difference-based convergence criteria, provides practical guidance for applying
Bayesian optimization to likelihood-based inference problems.

Future work includes extensions to multidimensional parameter spaces,
adaptive hyperparameter selection, and applications to nonlinear and
non-Gaussian state-space models, where exact likelihood evaluation is not
available.

\end{document}